\documentclass[letterpaper]{sig-alternate-2013}

\usepackage[bf]{caption}
\usepackage{times,eurosym}
\usepackage{subcaption}
\usepackage{paralist}
\usepackage{epsfig}
\usepackage[keeplastbox]{flushend}
\usepackage{enumitem}
\usepackage{algorithmic}
\usepackage{amssymb}
\usepackage{url}
\usepackage{algorithm}

\usepackage{color}
\usepackage{ifthen}
\newboolean{TECHREPORT}
\setboolean{TECHREPORT}{false}
\usepackage{cite}
\usepackage{amssymb,amsmath}
\usepackage{newclude}
\usepackage{epstopdf}
\usepackage{xspace}
\usepackage{graphicx}
\usepackage{url}
\usepackage{booktabs}
\usepackage[T1]{fontenc}
\usepackage{balance}
\usepackage{xspace}
\makeatletter
\def\url@leostyle{%
  \@ifundefined{selectfont}{\def\UrlFont{}}%
  {\def\UrlFont{}}%
}
\makeatother	
\usepackage[hyphenbreaks]{breakurl}
\newcommand{\descr}[1]{\smallskip\noindent\textbf{#1}}
\newcommand{\descrit}[1]{\smallskip\noindent\textit{#1}}
\newcommand{\hasht}[1]{{\sf{\small \##1}}\xspace}
\newcommand{\dspol}[1]{{\sf{\//pol\//}}\xspace}
\newcommand{\dsrk}[1]{{\sf{\//r9k\//}}\xspace}

\definecolor{darkgreen}{RGB}{47,109,79}
\definecolor{darkblue}{RGB}{57,79,99}
\newif\ifcomment
\commenttrue 
\newif\ifwatermark
\watermarkfalse

\ifwatermark
      \usepackage{draftwatermark}
      \SetWatermarkScale{.7}
      \SetWatermarkText{\shortstack{In Submission:\\Do Not Distribute}}
      \SetWatermarkColor[rgb]{1,0.88,0.88}
\fi

\permission{\copyright 2017 International World Wide Web Conference Committee (IW3C2),\\ published under Creative Commons CC BY 4.0 License.}
\conferenceinfo{WWW 2017 Companion}{April 3--7, 2017, Perth, Australia.}
\copyrightetc{ACM \the\acmcopyr}
\crdata{978-1-4503-4914-7/17/04. \\
http://dx.doi.org/10.1145/3041021.3053890 \\
\includegraphics[scale=0.5]{cc.png}}

\makeatletter
\def\@copyrightspace{\relax}
\makeatother
\begin{document} 

\newenvironment {squishlist}
{\begin{list}{$\bullet$}
  { \setlength{\itemsep}{1pt}
     \setlength{\parsep}{1pt}
     \setlength{\topsep}{1pt}
     \setlength{\partopsep}{1pt}
     \setlength{\leftmargin}{1.5em}
     \setlength{\labelwidth}{1em}
     \setlength{\labelsep}{0.5em} } }
{\end{list}}

\title{Measuring \#GamerGate:\\ A Tale of Hate, Sexism, and Bullying}

\numberofauthors{1}
\author{
Despoina Chatzakou$^{\dagger}$, Nicolas Kourtellis$^{\ddagger}$, Jeremy Blackburn$^{\ddagger}$\\[0.5ex]
Emiliano De Cristofaro$^{\sharp}$, Gianluca Stringhini$^{\sharp}$, Athena Vakali$^{\dagger}$\\[1ex]
  \affaddr{$^\dagger$Aristotle University of Thessaloniki \hspace{0.2cm} $^\ddagger$Telefonica Research 
  \hspace{0.2cm} $^{\sharp}$University College London}\\
 \affaddr{deppych@csd.auth.gr, nicolas.kourtellis@telefonica.com, jeremy.blackburn@telefonica.com}\\  \affaddr{e.decristofaro@ucl.ac.uk, g.stringhini@ucl.ac.uk, avakali@csd.auth.gr}
 \vspace{0.3cm}}

\maketitle

\begin{abstract}
Over the past few years, online aggression and abusive behaviors have occurred in many different forms and on a variety of platforms. 
In extreme cases, these incidents have evolved into hate, discrimination, and bullying, and even materialized into real-world threats and attacks against individuals or groups. 
In this paper, we study the Gamergate controversy. Started in August 2014 in the online gaming world, it quickly spread across various social networking platforms, ultimately leading to many incidents of cyberbullying and cyberaggression. 
We focus on Twitter, presenting a measurement study of a dataset of 340k unique users and 1.6M tweets to study the properties of these users, the content they post, and how they differ from random Twitter users. 
We find that users involved in this ``Twitter war'' tend to have more friends and followers, are generally more engaged and post tweets with negative sentiment, less joy, and more hate than random users. 
We also perform preliminary measurements on how the Twitter suspension mechanism deals with such abusive behaviors. 
While we focus on Gamergate, our methodology to collect and analyze tweets related to aggressive and bullying activities is of independent interest.
\end{abstract}

\section{Introduction}\label{sec:intro}

With the proliferation of social networking services and always-on always-connected devices, social interactions have increasingly moved online, as social media has become an integral part of people's every day life.
At the same time, however, new instantiations of negative interactions have arisen, including aggressive and bullying behavior among online users.
Cyberbullying, the digital manifestation of bullying and aggressiveness in online social interactions, has spread to various platforms such as Twitter~\cite{sanchez2011twitter}, Youtube~\cite{Chen2012DetectingOffensiveLanguage}, Ask.fm~\cite{Hosseinmardi2014TowardsUnderstandingCyberbullying,Hosseinmardi2015}, and Facebook~\cite{Hee2015AutomaticDetectionPreventionCyberbullying}.
Other community-based services such as Yahoo Answers~\cite{kayes2015ya-abuse}, and online gaming platforms are not an exception~\cite{Mortensen2016}. 
Research has showed that bullying actions are often organized, with online users called to participate in hateful raids against other social network users~\cite{hine2016longitudinal}.

The Gamergate controversy~\cite{Massanari09102015} is one example of a coordinated campaign of harassment in the online world.
It started with a blog post by an ex-boyfriend of independent game developer Zoe Quinn, alleging sexual improprieties.
4chan boards like \dsrk~\cite{r9kthread} and \dspol~\cite{polthread}, turned it into a narrative about ``ethical'' concerns in video game journalism and began organizing harassment campaigns~\cite{hine2016longitudinal}.
It quickly devolved into a polarizing issue, involving sexism, feminism, and ``social justice,'' taking place on social media like Twitter~\cite{guberman2017}.
Although held up as a pseudo-political movement by its adherents, there is substantial evidence that Gamergate is more accurately described as an organized campaign of hate and harassment~\cite{theguardianfeliciaday}.
What started as ``mere'' denigration of women in the gaming industry, eventually evolved into directed threats of violence, rape, and murder~\cite{nytimessarkeesiandeaththreats}.
Gamergate came about due to a unique time in the digital world in general, and gaming in particular.
The recent democratization of video game development and distribution via platforms such as Steam has allowed for a new generation of ``indie'' game developers who often have a more intimate relationship with their games and the community of gamers that play them.
With the advent of ubiquitous social media and a community born in the digital world, the Gamergate controversy provides us a unique point of view into online harassment campaigns.

\descr{Roadmap.} In this paper, we explore a slice of the Gamergate controversy by analyizing  1.6M tweets from 340k unique users part of whom engaged in it.
As a first attempt at quantifying this controversy, we focus on how these users, and the content they post, differ from random (baseline) Twitter users.
We discover that Gamergaters are seemingly more engaged than random Twitter users, which is an indication as to how and why this controversy is still on going.
We also find that, while their tweets appear to be aggressive and hateful, Gamergaters do \emph{not} exhibit common expressions of online anger, and in fact primarily differ from random users in that their tweets are less joyful.
The increased rate of engagement of Gamergate users makes it more difficult for Twitter to deal with all these cases at once, something reflected in the relative low suspension rates of such users.
In the struggle to combat existing aggressive and bullying behaviors, Twitter recently took new actions and is now temporarily limiting users for abusive behavior~\cite{independent}. 

Finally, we note that, although our work is focused on Gamergate in particular, our principled methodology to collect and analyze tweets related to aggressive and bullying activities on Twitter can be generalized and it is thus of independent interest.

\descr{Paper Organization.} Next section reviews related work, then Section~\ref{sec:methodology} discusses our data collection methodology.
In Section~\ref{sec:results}, we present the results of our analysis and lessons we learn from them.
Finally, the paper concludes in Section~\ref{sec:discussion}.

\section{Related work}\label{sec:related-work}

Previous research has studied and aimed at detecting offensive, abusive, aggressive, and bullying content on social media, including Twitter, YouTube, Instagram, Facebook and Ask.fm.
Next, we cover related work on this type of behavior in general, as well as work related to the Gamergate case.

\descr{Detecting abusive behavior.} 
Chen et al.~\cite{Chen2012DetectingOffensiveLanguage} aim to detect offensive content and potential offensive users by analyzing YouTube comments.
Then, Hosseinmardi et al.~\cite{Hosseinmardi2014TowardsUnderstandingCyberbullying,Hosseinmardi2015} turn to cyberbullying on Instagram and Ask.fm.
Specifically, in~\cite{Hosseinmardi2015}, besides considering available text information, they also try to associate the topic of an image (e.g., drugs, celebrity, sports, etc.) to possible cyberbullying events, concluding that drugs are highly associated with cyberbullying.
Also, in a effort to create a suitable dataset for their analysis, at first the authors collected a large number of media sessions -- i.e., videos and images along with comments -- from Instagram public profiles, with a subset selected for labeling.
To ensure that an adequate number of cyberbullying instances will be present in the dataset, they selected media sessions with at least one profanity word.
Finally, they relied on the CrowdFlower crowdsourcing platform to determine whether or not such sessions are related with cyberbullying or cyberaggression.
In~\cite{Hosseinmardi2014TowardsUnderstandingCyberbullying} authors leveraged both likes and comments to identify negative behavior in the Ask.fm social network.
Here, their dataset was created by exploiting publicly accessible profiles, e.g. questions, answers, and likes.

Other works aim to detect hate/abusive content on Yahoo Finance.
In~\cite{Djuric2015HateSpeechDetection}, the authors use a Yahoo Finance dataset labeled over a 6-month period.
Nobata et al.~\cite{Nobata2016AbusiveLanguageDetection} gather a new dataset from Yahoo Finance and News comments: each comment is initially characterized as either \textit{abusive} or \textit{clean} (from Yahoo's in-house trained raters), with further analysis on the abusive comments specifying whether they contain hate, derogative language, or profanity.
They follow two annotation processes, with labeling performed by: (i)~three trained raters, and (ii)~workers recruited from Amazon Mechanical Turk, concluding that the former is more effective.

Kayes et al.~\cite{kayes2015ya-abuse} focus on a Community-based Question-Answering (CQA) site, Yahoo Answers, finding that users tend to flag abusive content posted in an overwhelmingly correct way, while in~\cite{Dinakar2011ModelingDetectionTextualCyberbullying}, the problem of cyberbullying is further decomposed to sensitive topics related to race and culture, sexuality, and intelligence, using YouTube comments extracted from controversial videos.
Hee et al.~\cite{Hee2015AutomaticDetectionPreventionCyberbullying} also study specific types of cyberbullying, e.g., threats and insults, on Dutch posts extracted from Ask.fm social media.
They also highlight three main user behaviors, harasser, victim, and bystander -- either bystander-defender or bystander-assistant who support the victim or the harasser, respectively.
Their dataset was created by crawling a number of seed sites from Ask.fm, with a limited number of cyberbullying instances. 
They complement the data with more cyberbullying related content by: (i)~launching a campaign where people reported personal cases of cyberbullying taking place in different platforms, i.e., Facebook, message board posts and chats, and (ii)~by designing a role-playing game involving a cyberbullying simulation on Facebook.
Then, they ask manual annotators to characterize content as being part of a cyberbullying event, and indicate the author's role in such event, i.e., victim, harasser, bystander-defender, or bystander-assistant.

Sanchez et al.~\cite{sanchez2011twitter} use Twitter messages to detect bullying instances and more specifically cases related to gender bullying.
They use a distant supervision approach~\cite{BhayaniHuang2009} to automatically label a set of tweets by using a set of abusive terms used to characterize text as expressing negative or positive sentiment.
The dataset is then used to train a classifier geared to finding inappropriate words in Twitter text and  detect bullying -- the hypothesis being that bullying instances most probably contain negative sentiment.
Finally, in~\cite{chatzakou2017icwsm} the authors propose an approach suitable for detecting bullying and aggressive behavior on Twitter.
They study the properties of cuberbullies and aggressors and what distinguishes them from regular users.
To perform their analysis, they build upon the CrowdFlower crowdsourcing tool to create a dataset where users are characterized as bullies, aggressors, spammers, or normal.

Even though Twitter is among the most popular social networks, only a few efforts have focused on detecting abusive content on it.
Here, we propose an approach for building a ground truth dataset, using Twitter as a source of information, which will contain a higher density of abusive content (mimicking real life abusive posting activity).

\descr{Analysis of Gamergate.} In our work, the hashtag \hasht{GamerGate} serves as a seed word to build a dataset of abusive behavior, as Gamergate is one of the best documented large-scale instances of bullying/aggressive behavior we are aware of~\cite{Massanari09102015}. 
With individuals on both sides of the controversy using it, and extreme cases of cyberbullying and aggressive behavior associated with it (e.g., direct threats of rape and murder), \hasht{GamerGate} is a relatively unambiguous hashtag associated with tweets that are likely to involve abusive/aggressive behavior. Prior work has also looked at Gamergate in somewhat related contexts.
For instance, Guberman and Hemphill~\cite{guberman2017} used \hasht{GamerGate} to collect a sufficient number of harassment-related tweets in an effort to study and detect toxicity on Twitter.
Also, Mortensen~\cite{Mortensen2016} likens the Gamergate phenomenon to hooliganism, i.e., a leisure-centered aggression where fans are organized in groups to attack another group's members.

\section{Methodology}\label{sec:methodology}

In this section, we present our methodology for collecting and processing a dataset
of abusive behavior on Twitter. In this paper, we focus on the Gamergate case, however, our methodology can be generalized to other platforms and case studies.

\subsection{Data Collection}

\descr{Seed keyword(s).}
The first step is to select one or more seed keywords, which are likely related to the occurrence of abusive incidents.
Besides \hasht{GamerGate}, good examples are also \hasht{BlackLivesMatter} and \hasht{PizzaGate}.
In addition to such seed words, a set of hate- or curse-related words can also be used, e.g., words extracted from the Hatebase database (HB)\footnote{ \url{https://www.hatebase.org/}}, to start collecting possible abusive texts from social media sources.
Therefore, at time $t_1$, the list of words to be used for filtering posted texts includes only the seed word(s), i.e., $L(t_1) = <seed(s)>$.

In our case, we focus on Twitter, and more specifically we build upon the Twitter Streaming API\footnote{ \url{https://dev.twitter.com/streaming/overview}}
which gives access to 1\% of all tweets.
This returns a set of correlated information, either user-based, e.g., poster's username, followers and friends count, profile image, total number of posted/liked/favorite tweets, or text-based, e.g., the text itself, hashtags, URLs, if it is a retweeted or reply tweet, etc.
The data collection process took place from June to August 2016.
Initially, we obtained a 1\% of the sample public tweets and parsed it to select all tweets containing the seed word \hasht{GamerGate}, which are likely to involve the type of behavior, and the case study we are interested in.

\descr{Dynamic list of keywords.}
In addition to the seed keyword(s), further filtering keywords are used to select abusive-related content.
The list of the additional keywords is updated dynamically in consecutive time intervals based on the posted texts during these intervals.
Thus, in $T = \{t_1, t_2, ..., t_n\}$ the keywords list L(t) has the following form:
$L(t_i) = <seed(s), kw_1, kw_2€, kw_N>$, where $kw_j$ is the $j$th top keyword in time period $\Delta T = t_i - t_{i-1}$.
Depending on the topic under examination, i.e., if it is a popular topic or not, the creation of the dynamic keywords list can be split to different consecutive time intervals.
To maintain the dynamic list of keywords for the time period $t_{i-1} \rightarrow t_{i}$, we investigate the texts posted in this time period.
We extract $N$ keywords found during that time, compute their frequency and rank them into a temporary list $LT(t_i)$.
We then adjust the dynamic list $L(t_i)$ with entries from the temporary list $LT(t_i)$ to create a new dynamic list that contains the up-to-date top N keywords along with the seed words.
This new list is used in the next time period $t_{i} \rightarrow t_{i+1}$ for the filtering of posted text.

As mentioned, \hasht{GamerGate} serves as a seed for a snowball sampling of other hashtags likely associated with cyberbullying and aggressive behavior.
We include tweets with hashtags appearing in the same tweets as \hasht{GamerGate} (the keywords list is updated on a daily basis).
Overall, we reach 308 hashtags during the data collection period.
A manual examination of these hashtags reveals that they do contain a number of hate words, e.g., \hasht{InternationalOffendAFeministDay}, \hasht{IStandWithHateSpeech}, and \hasht{KillAllNiggers}.

\descr{Random sample.}
To complement the dataset with cases that are less likely to contain abusive content, we also crawl a random sample of texts over the same time period.
In our case, we simply crawl a random set of tweets, which constitutes our baseline.

\descr{Remarks.} Overall, we have collected two datasets: (i) a random sample set of 1M tweets, and (ii) a set of 659k tweets which are likely to contain abusive behavior.

We argue that the our data collection methodology provides several benefits with respect to performance.
First, it allows for regular updates of the keyword list, hence, the collection of more up-to-date content and capturing previously unseen behaviors, keywords, and trends.
Second, it lets us adjust the update time of the dynamic keywords list based on the observed burstiness of the topics under examination, thus eliminating the possibility of either losing new information or collecting repeatedly the same information.
Finally, this process can be parallelized for scalability on multiple machines using a Map-Reduce script for computing top N keywords list.
All machines maintain local top N keyword lists which are aggregated globally in a central controller, enabling the construction of a global dynamic top N keyword list that can be distributed back to the computing / crawling machines.

\subsection{Data Processing}

We performed preprocessing of the data collected to produce a `clean' dataset, free of noisy data. 

\descr{Cleaning.} We remove stop words, URLs, numbers, and punctuation marks.
Additionally, we perform normalization, i.e., we eliminate repeated letters and repetitive characters which users often use to express their feelings more intensely (e.g., the word `hellooo' is converted to `hello').

\descr{Spam removal.} Even though extensive work has been done on spam detection in social media, e.g.,~\cite{GiatsoglouNDSYNC,stringhini2010detecting,Wang2010SpamDetectionTwitter}, Twitter is still full of spam accounts~\cite{Chen2015SpamTweets}, often using vulgar language and exhibiting behavior (repeated posts with similar content, mentions, or hashtags) that could also be considered as aggressive or bullying.
So, to eliminate part of this noise we proceeded with a first-level spam removal process by considering two attributes which have already been used as filters (e.g.,~\cite{Wang2010SpamDetectionTwitter}) to remove spam incidents: (i) the number of hashtags per tweet (often used for boosting the visibility of the posted tweets), and (ii) posting of (almost) similar tweets.
To find optimal cutoffs for these heuristics, we study both the distribution of hashtags and the duplication of tweets. 

\descrit{Hashtags.} The hashtags distribution shows that users tend to use from 0 to about 17 hashtags on average.
With such information at hand, we test different cutoffs to set a proper limit, upon which the user could be characterized as spammer.
After a manual inspection on a sample of posts, we set the limit to 5 hashtags, i.e., users with more than 5 hashtags, on average, are flagged as spammers, and their tweets are removed from the dataset.

\begin{figure}[!t]
	\centering
	\includegraphics[width=0.27\textwidth]{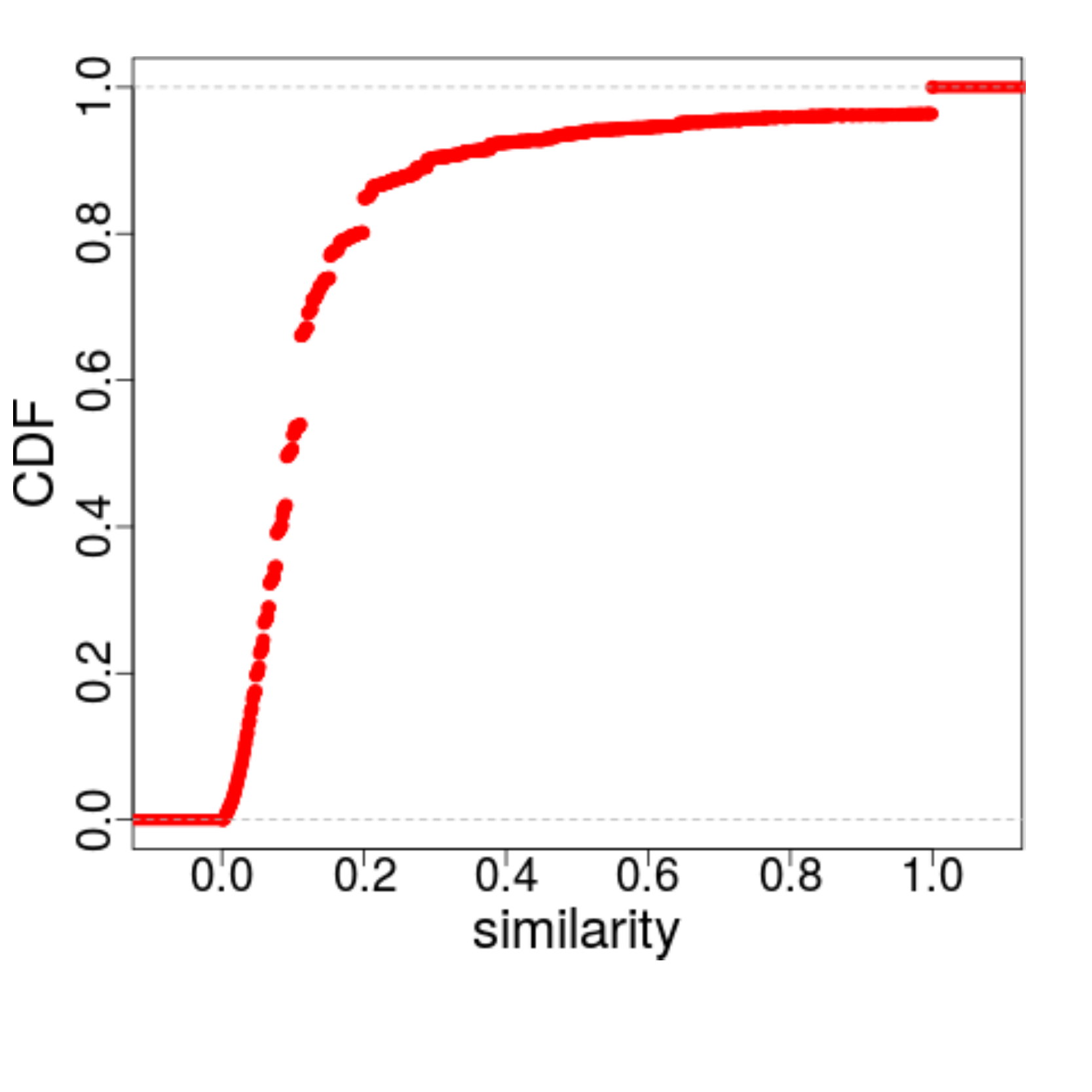}
	\vspace{-0.5cm}
	\caption{Similarity distribution.}
	\label{fig:duplications}
\end{figure}

\descrit{Duplications.} We also estimate the similarity of users' posts based on an appropriate similarity measure.
In many cases, a user's tweets are (almost) the same, while only the listed mentioned users are modified.
So, in addition to the previous presented cleaning processes, we also remove all existing mentions.
We then proceed to compute the Levenshtein distance~\cite{Navarro2001ApproximateStringMatching}, which counts the minimum number of single-character edits needed to convert one string into another, averaging it out over all pairs of their tweets.
Initially, for each user, we calculated the intra-tweet similarity, then we set to out estimate  the average intra-tweets similarity.
For a user with $x$ tweets, we use a set of $n$ similarity scores, where $n = x (x - 1) / 2$.
In the end, all users with intra-tweet similarity above $0.8$ are excluded from the dataset.
Figure~\ref{fig:duplications} shows that about 5\% of the users have a high percentage of similar posts and which were removed.

\begin{figure*}[!t]
	\centering
	\begin{subfigure}[b]{0.24\textwidth}
		\includegraphics[width=\textwidth]{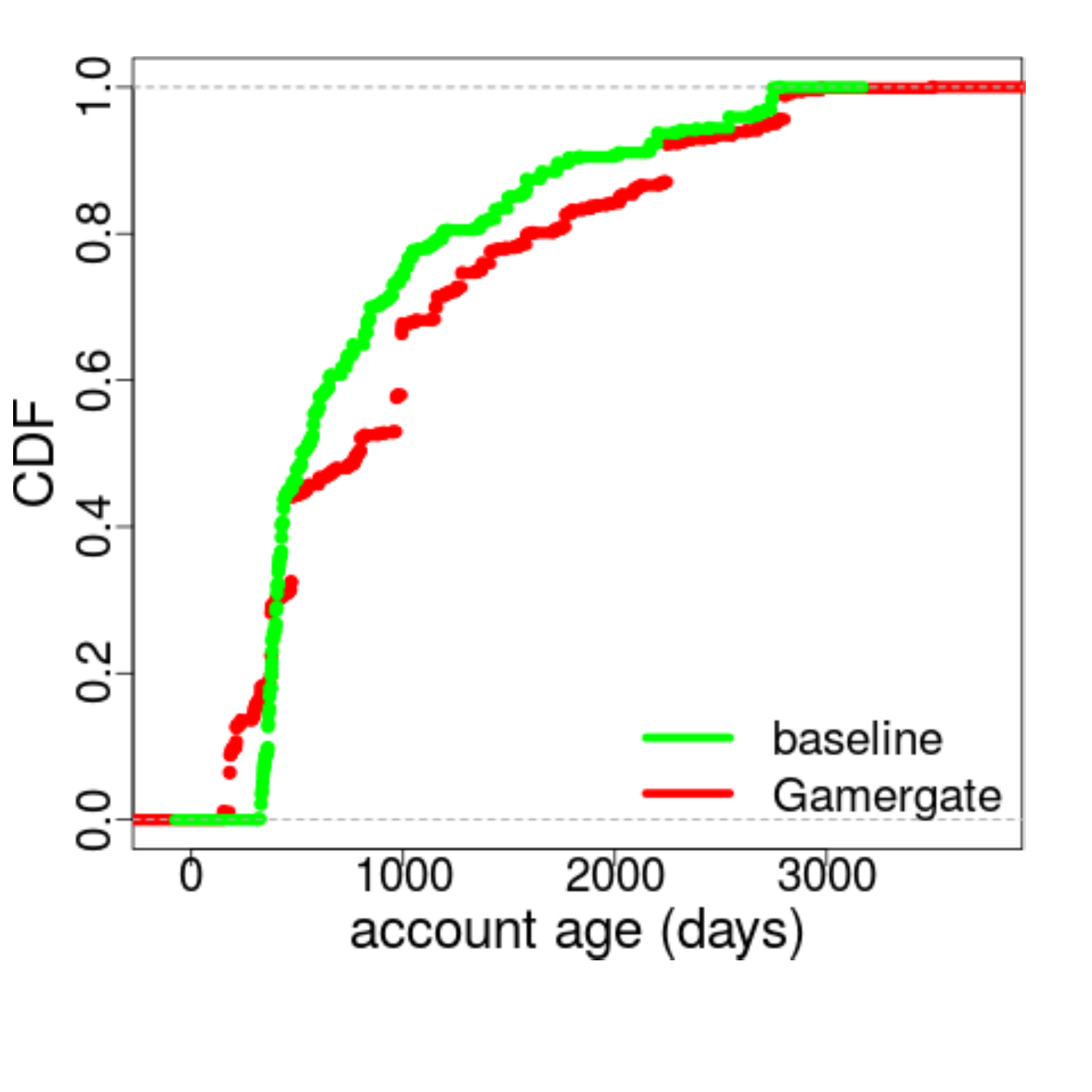}
		\captionsetup{font=scriptsize}
		\vspace{-0.6cm}
		\caption{Account age distribution.}
		\label{fig:baseline_hatebase_age}
	\end{subfigure}
	\begin{subfigure}[b]{0.24\textwidth}
		\includegraphics[width=\textwidth]{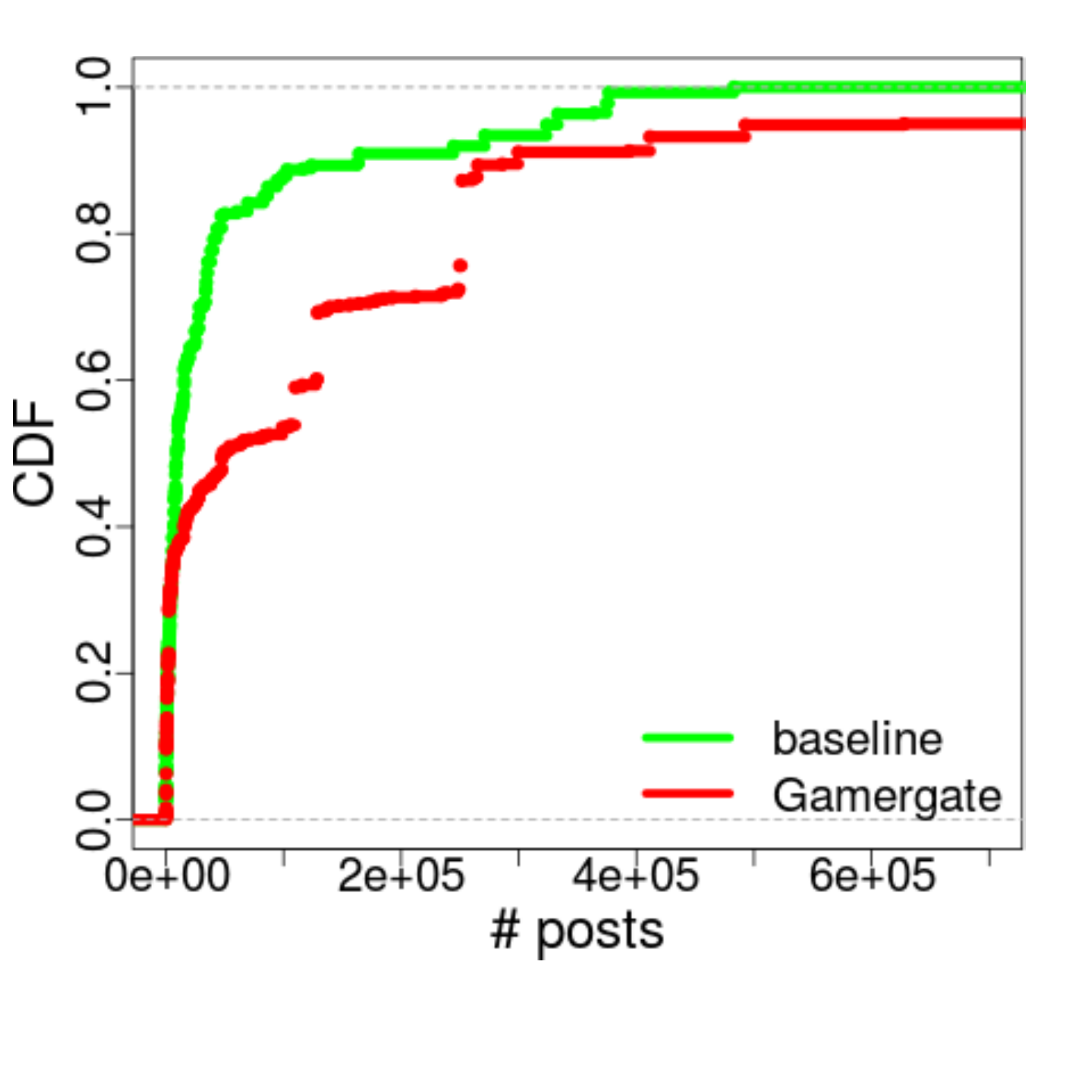}
		\captionsetup{font=scriptsize}
		\vspace{-0.6cm}
		\caption{Posts distribution.}
		\label{fig:baseline_hatebase_posts}
	\end{subfigure}
	\begin{subfigure}[b]{0.24\textwidth}
		\includegraphics[width=\textwidth]{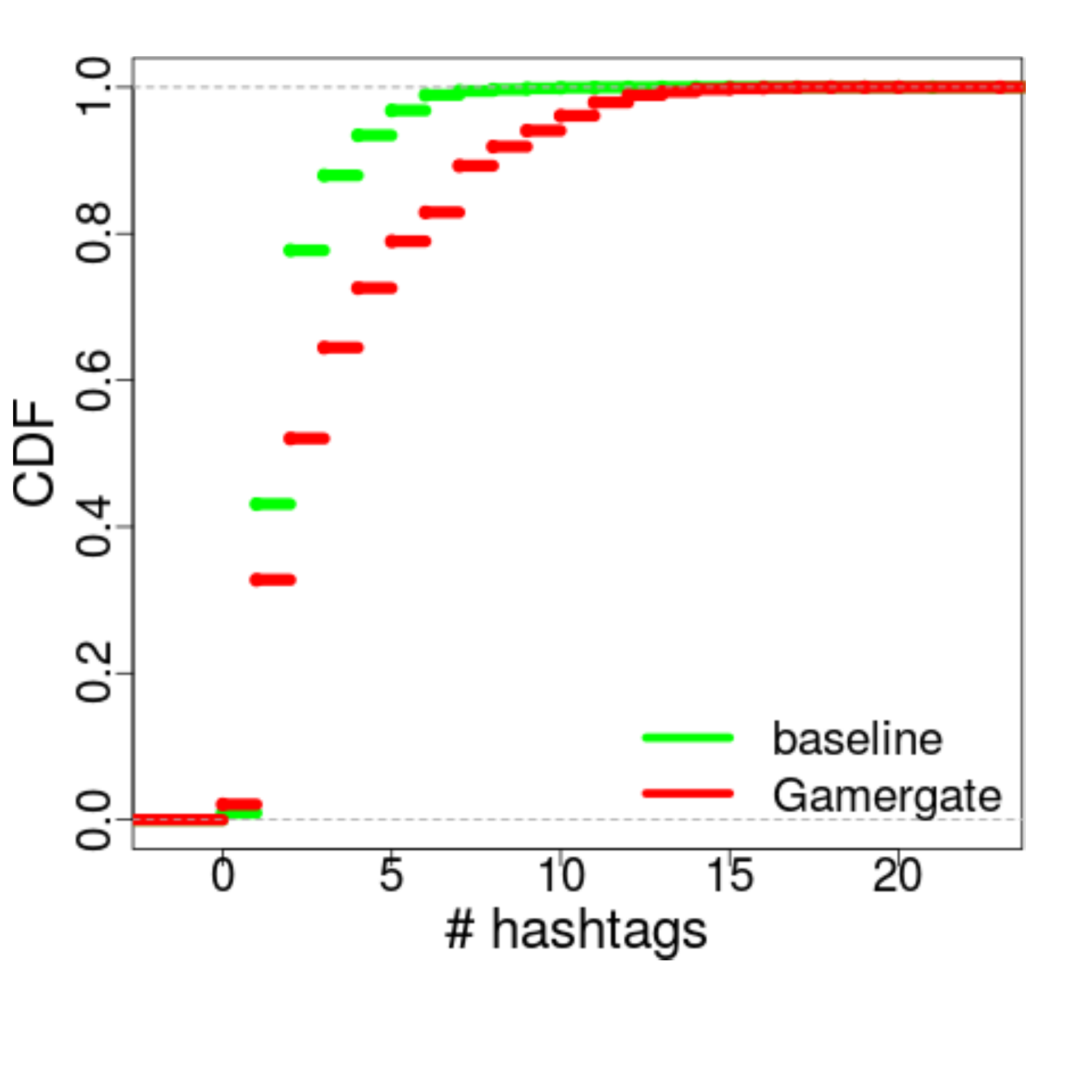}
		\captionsetup{font=scriptsize}
		\vspace{-0.6cm}
		\caption{Hashtags distribution.}
		\label{fig:baseline_hatebase_hashtags}
	\end{subfigure}
	\caption{ CDF of (a) Account age, (b) Number of posts, and (c) Hashtags.}
\end{figure*}

\begin{figure*}[!t]
	\centering
	\begin{subfigure}[b]{0.24\textwidth}
		\includegraphics[width=\textwidth]{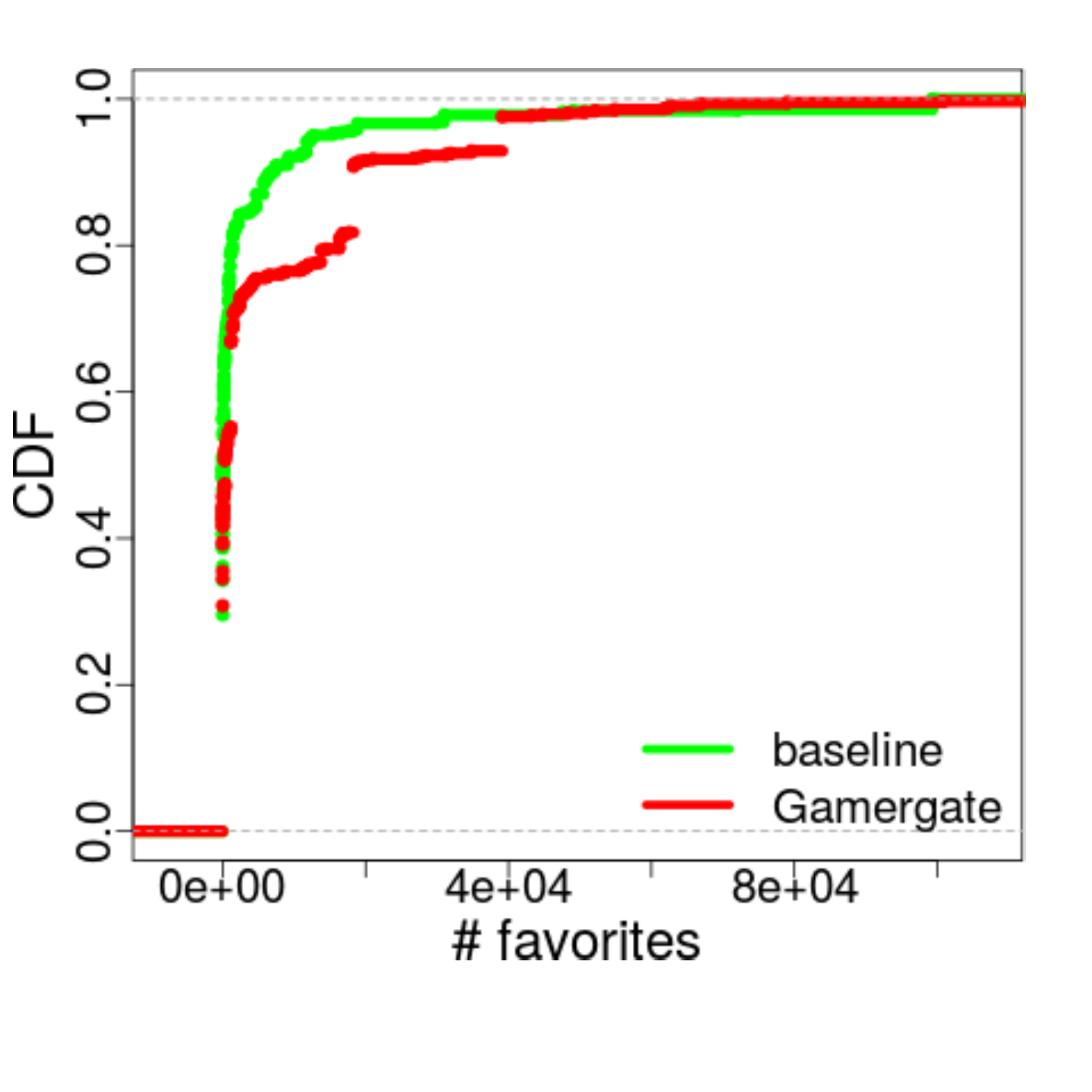}
		\captionsetup{font=scriptsize}
		\vspace{-0.6cm}
		\caption{Favorites distribution.}
		\label{fig:baseline_hatebase_favourites}
	\end{subfigure}
	\begin{subfigure}[b]{0.24\textwidth}
		\includegraphics[width=\textwidth]{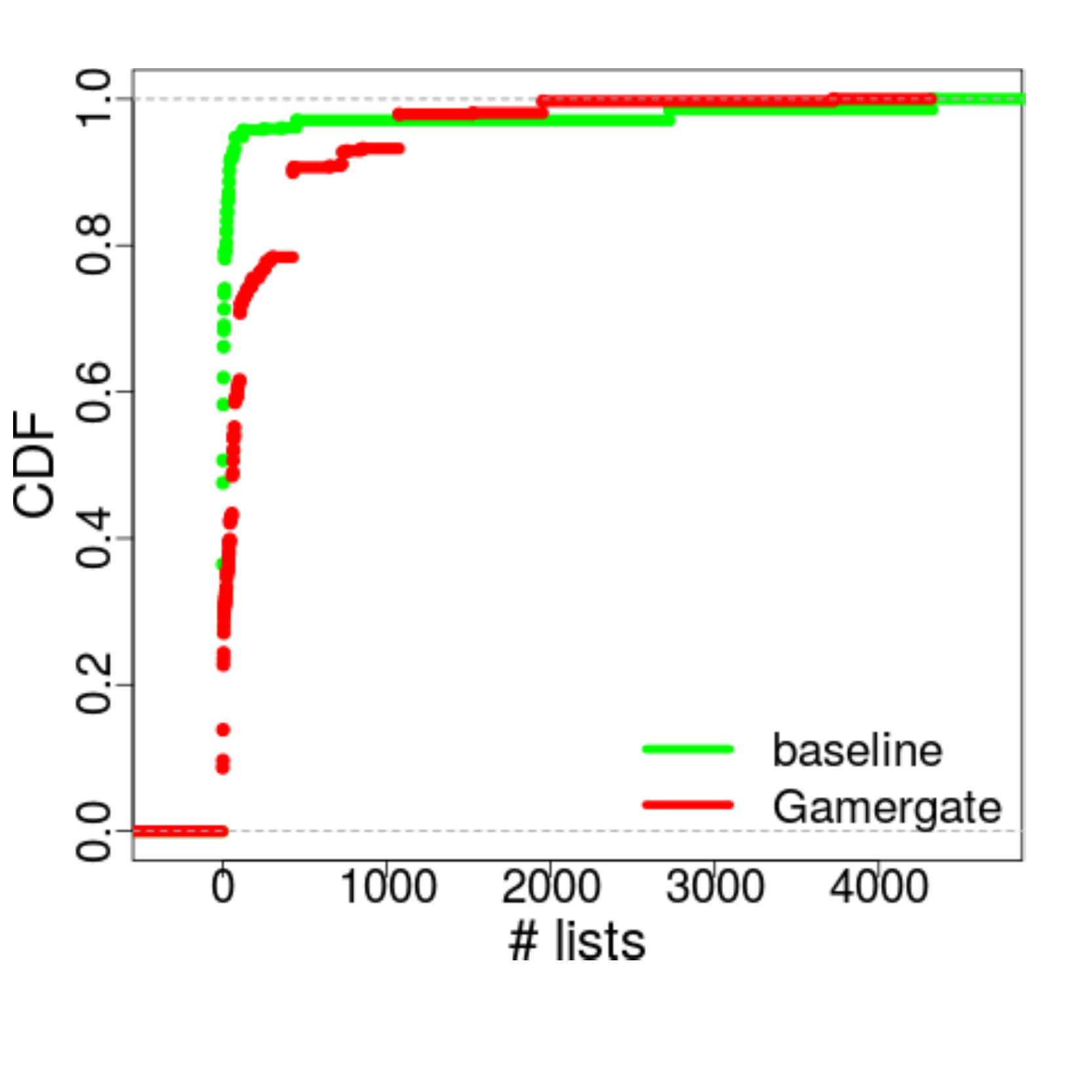}
		\captionsetup{font=scriptsize}
		\vspace{-0.6cm}
		\caption{Lists distribution.}
		\label{fig:baseline_hatebase_lists}
	\end{subfigure}
	\begin{subfigure}[b]{0.24\textwidth}
		\includegraphics[width=\textwidth]{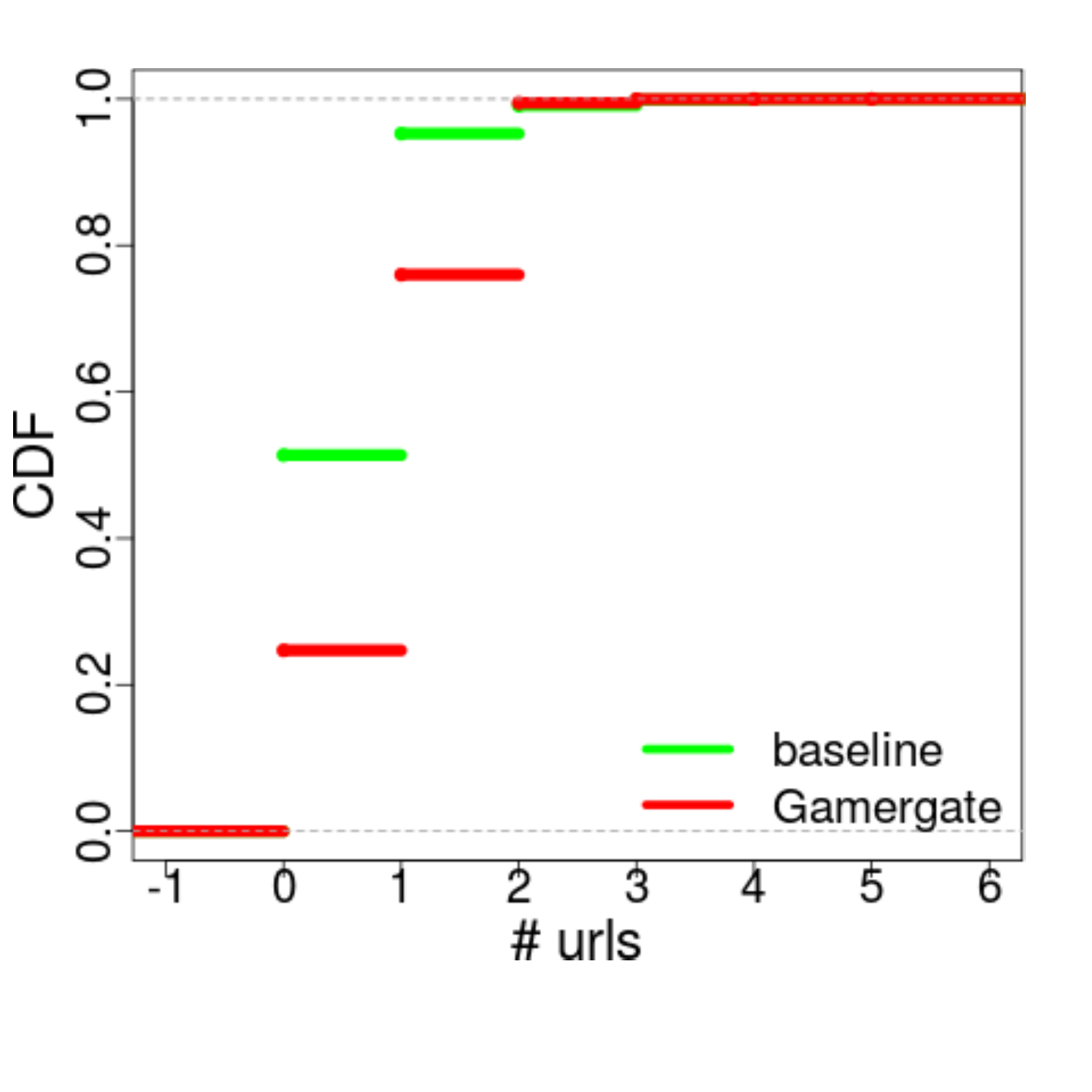}
		\captionsetup{font=scriptsize}
		\vspace{-0.6cm}
		\caption{URLs distribution.}
		\label{fig:baseline_hatebase_urls}
	\end{subfigure}
	\begin{subfigure}[b]{0.24\textwidth}
		\includegraphics[width=\textwidth]{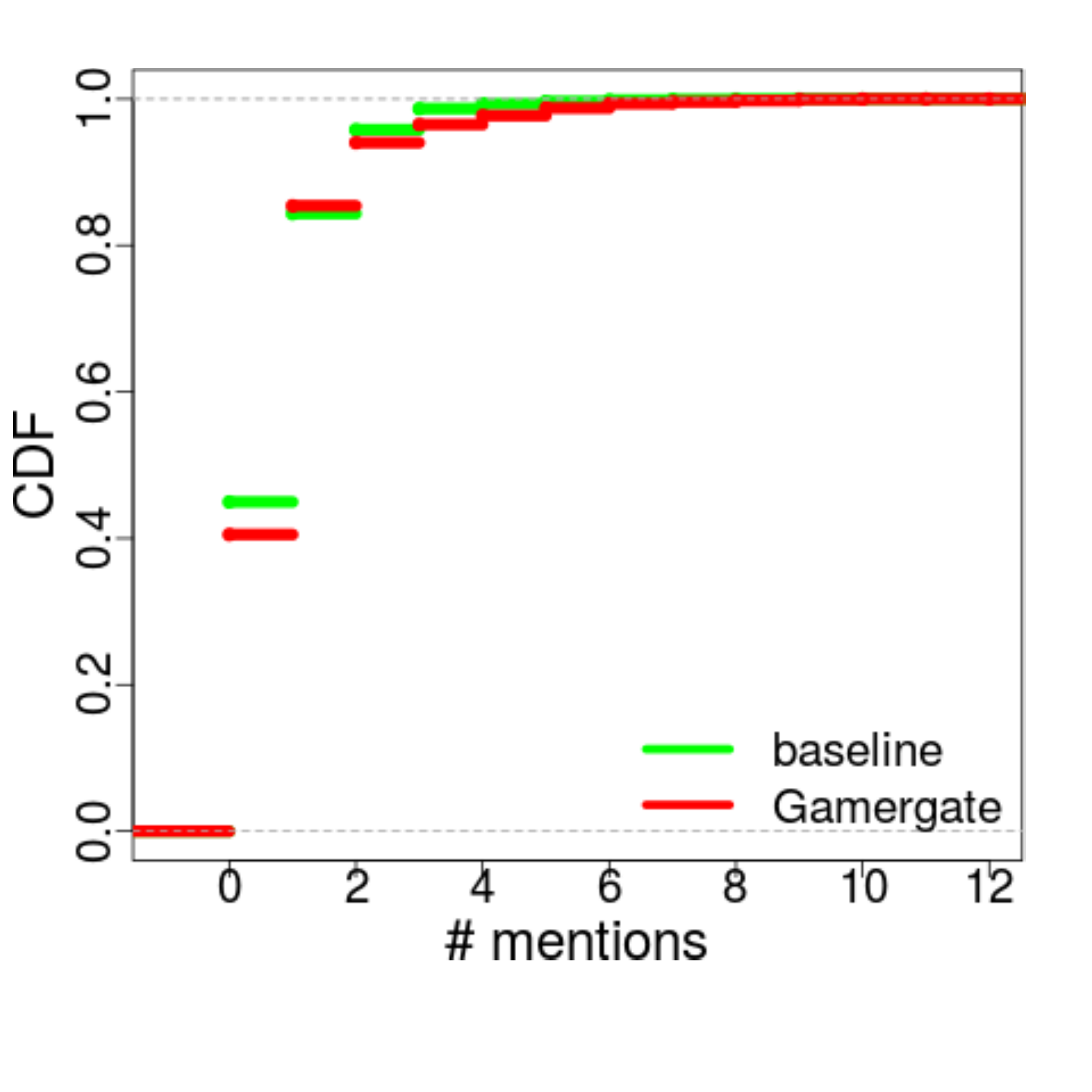}
		\captionsetup{font=scriptsize}
		\vspace{-0.6cm}
		\caption{Mentions distribution.}
		\label{fig:baseline_hatebase_mentions}
	\end{subfigure}	
		\vspace{-0.1cm}
	\caption{ CDF of (a) Number of Favorites, (b) Lists, (c) URLs, (d) Mentions.}
		\vspace{-0.1cm}

\end{figure*}

\section{Results} \label{sec:results}

In this section, we present the results of our measurement-based characterization, comparing the baseline and the Gamergate (GG) related datasets across various dimensions, including user attributes, posting activity, content semantics, and  Twitter account status.

\subsection{How Active are Gamergaters?}

\descr{Account age.}
An underlying question about the Gamergate controversy is what started first: participants tweeting about it or Twitter users participating in Gamergate? In other words, 
did Gamergate draw people to Twitter, or were Twitter users drawn to Gamergate?
In Figure~\ref{fig:baseline_hatebase_age}, we plot the distribution of account age for users in the Gamergate dataset and baseline users.
For the most part, GG users tend to have older accounts ($mean = 982.94$ days, $median = 788$ days, $STD = 772.49$ days).
The mean, median, and STD values for the random users are $834.39$, $522$, and $652.42$ days, respectively.
Based on a two-sample Kolmogorov-Smirnov test,\footnote{A statistical test to compare the probability distributions of different samples.} the two distributions are different with a test statistic $D = 0.20142$ and $p < 0.01$.
Overall, the oldest account in our dataset belongs to a GG user, while only $26.64\%$ of baseline accounts are older than the mean value of the GG users.
Figure~\ref{fig:baseline_hatebase_age} indicates that GG users were existing Twitter users drawn to the controversy.
In fact, their familiarity with Twitter could be the reason that Gamergate exploded in the first place.

\descr{Tweets and Hashtags.} In Figure~\ref{fig:baseline_hatebase_posts}, we plot the distribution of the number of tweets made by GG users and random users.
GG users are significantly more active than random Twitter users ($D = 0.352$, $p < 0.01$).
The mean, median, and STD values for the GG (random) users are 135,618 (49,342), 48,587 (9,429), and 185,997 (97,457) posts, resp.
Figure~\ref{fig:baseline_hatebase_hashtags} reports the CDF of the number of hashtags found in users' tweets for both GG and the random sample, finding that GG users use significantly ($D = 0.25681$, $p < 0.01$) more hashtags than random Twitter users.

\descr{Other characteristics.}
Figures~\ref{fig:baseline_hatebase_favourites} and~\ref{fig:baseline_hatebase_lists} show the CDFs of favorites and lists declared in the users' profiles.
We note that in the median case, GG users are similar to baseline users, but on the tail end (30\% of users), GG users have more favorites and topical lists declared than random users.
Then, Figure~\ref{fig:baseline_hatebase_urls} reports the CDF of the number of URLs found in tweets by both baseline and GG users.
The former post fewer URLs (the median indicates a difference of 1-2 URLs, $D = 0.26659$, $p < 0.01$), while the latter post more in an attempt to disseminate information about their ``cause,'' somewhat using Twitter like a news service.
Finally, Figure~\ref{fig:baseline_hatebase_mentions} shows that GG users tend to make more mentions within their posts, which can be ascribed to the higher number of direct attacks compared to random users.

\descr{Take aways.} Overall, the behavior we observe is indicative of GG users' ``mastery'' of Twitter as a mechanism for broadcasting their ideals. They make use of more advanced features, e.g., lists, tend to favorite more tweets, and share more URLs and hashtags than random users.
Using hashtags and mentions can draw attention to their message, thus GG users likely use them to disseminate their ideas deeper in the Twitter network, possibly aiming to attack more users and topical groups.

\begin{figure}[!t]
	\centering
	\begin{subfigure}[b]{0.23\textwidth}
		\includegraphics[width=\textwidth]{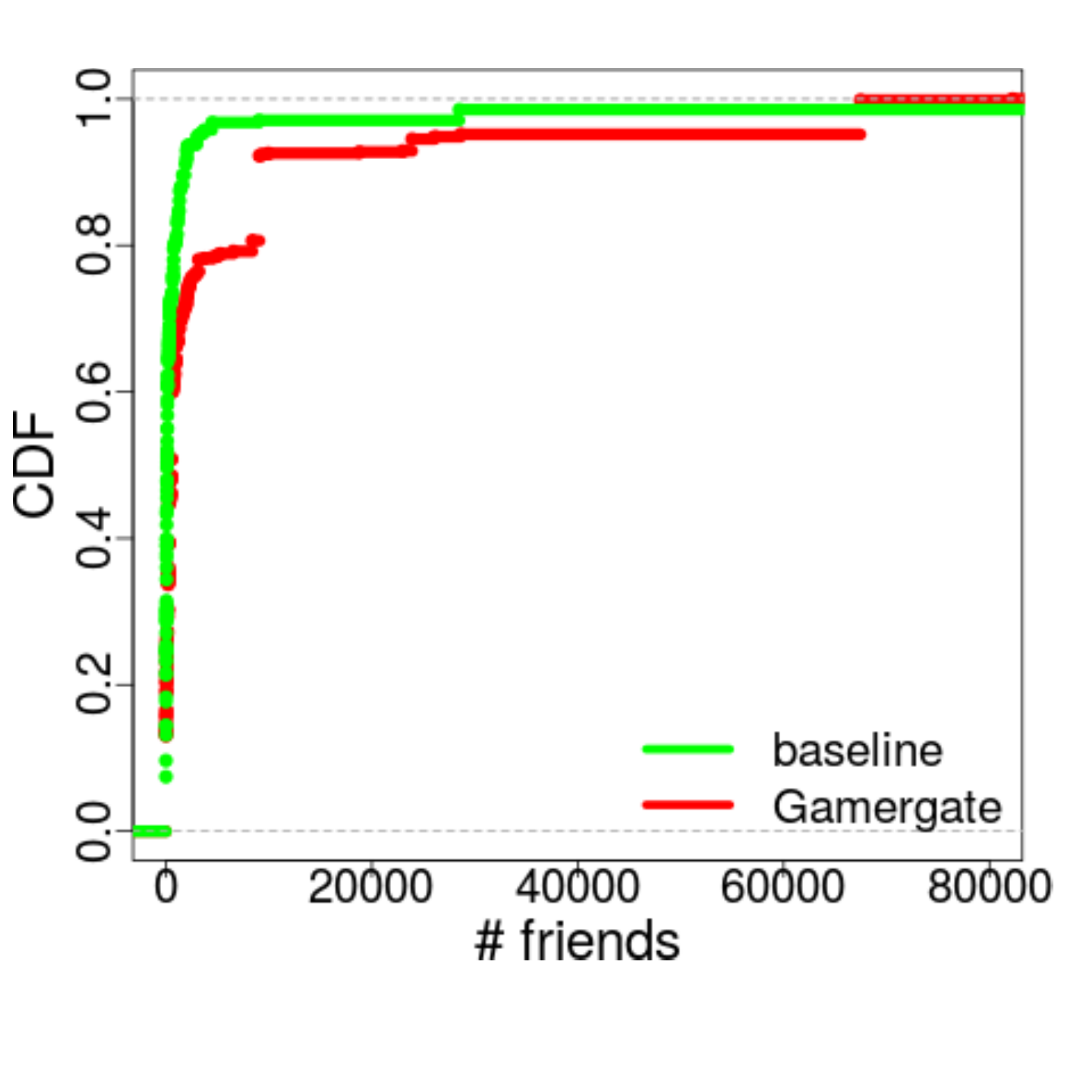}
		\captionsetup{font=scriptsize}
		\vspace{-0.6cm}
		\caption{Friends distribution.}
		\label{fig:baseline_hatebase_friends}
	\end{subfigure}
	\begin{subfigure}[b]{0.23\textwidth}
		\includegraphics[width=\textwidth]{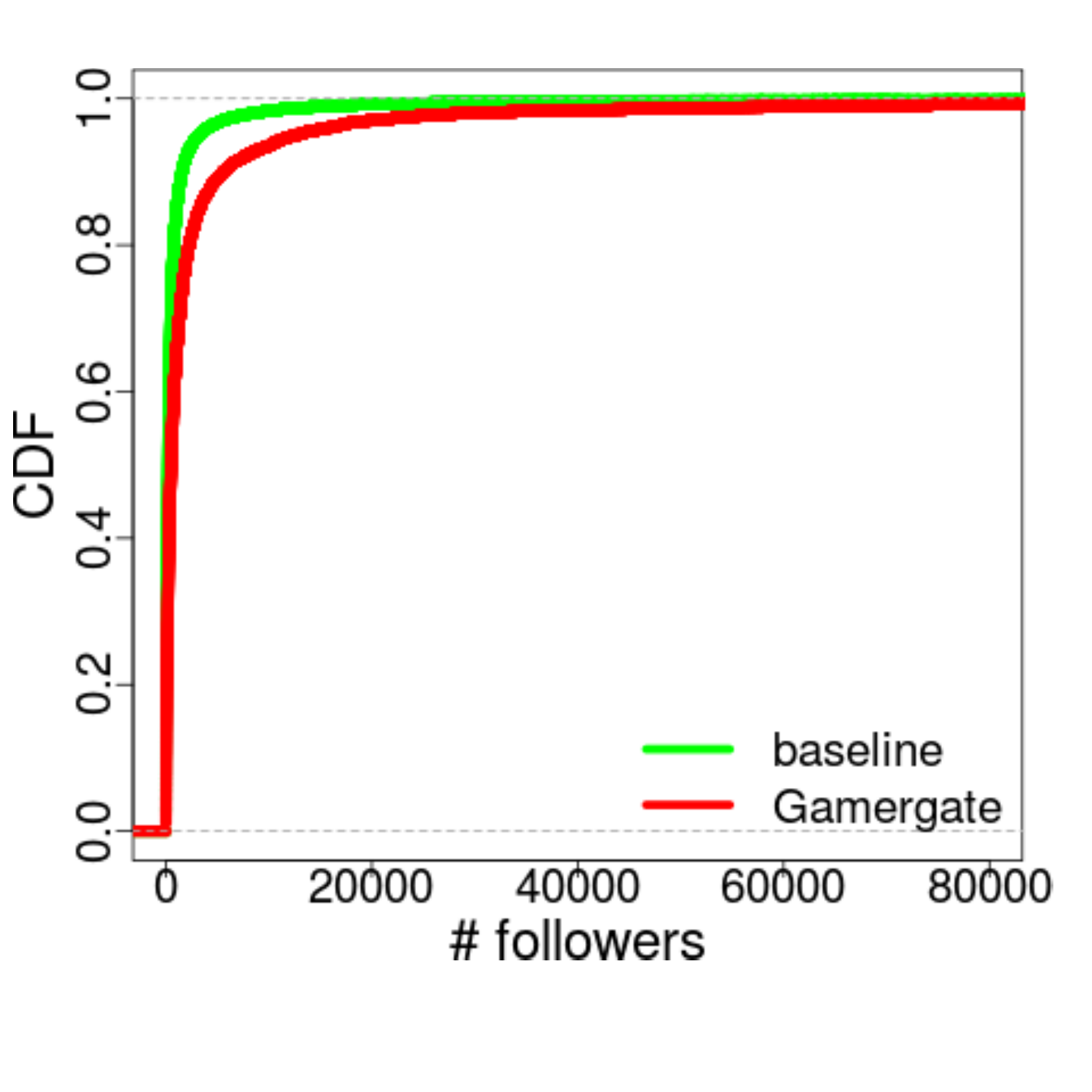}
		\captionsetup{font=scriptsize}
		\vspace{-0.6cm}
		\caption{Followers distribution.}
		\label{fig:baseline_hatebase_followers}
	\end{subfigure}
	\caption{ CDF of (a) Number of Friends, (b) Followers.}
\end{figure}

\begin{figure*}[!t]
	\centering
	\begin{subfigure}[b]{0.24\textwidth}
		\includegraphics[width=\textwidth]{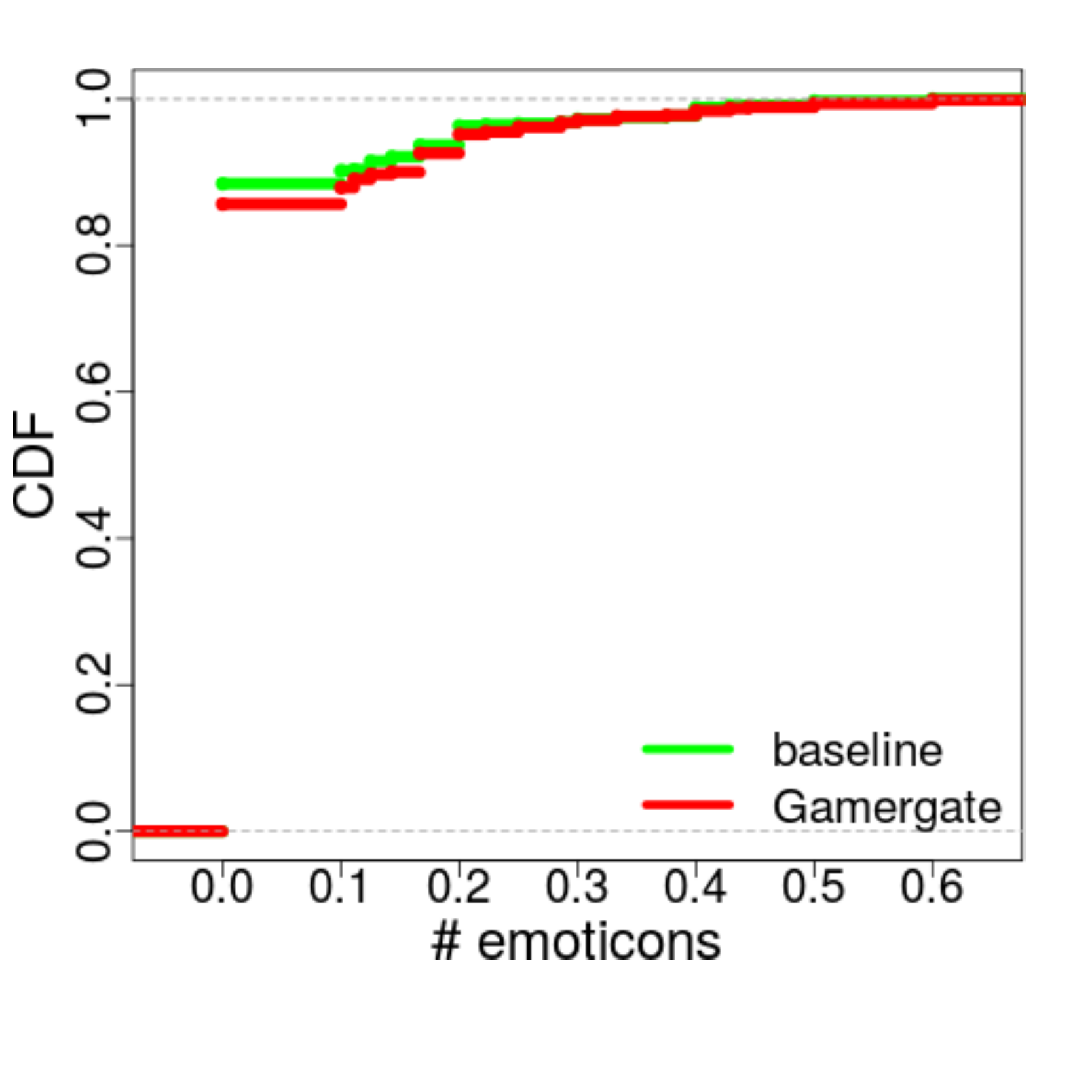}
		\captionsetup{font=scriptsize}
		\vspace{-0.6cm}
		\caption{Emoticons distribution.}
		\label{fig:baseline_hatebase_emoticons}
	\end{subfigure}
	\begin{subfigure}[b]{0.24\textwidth}
		\includegraphics[width=\textwidth]{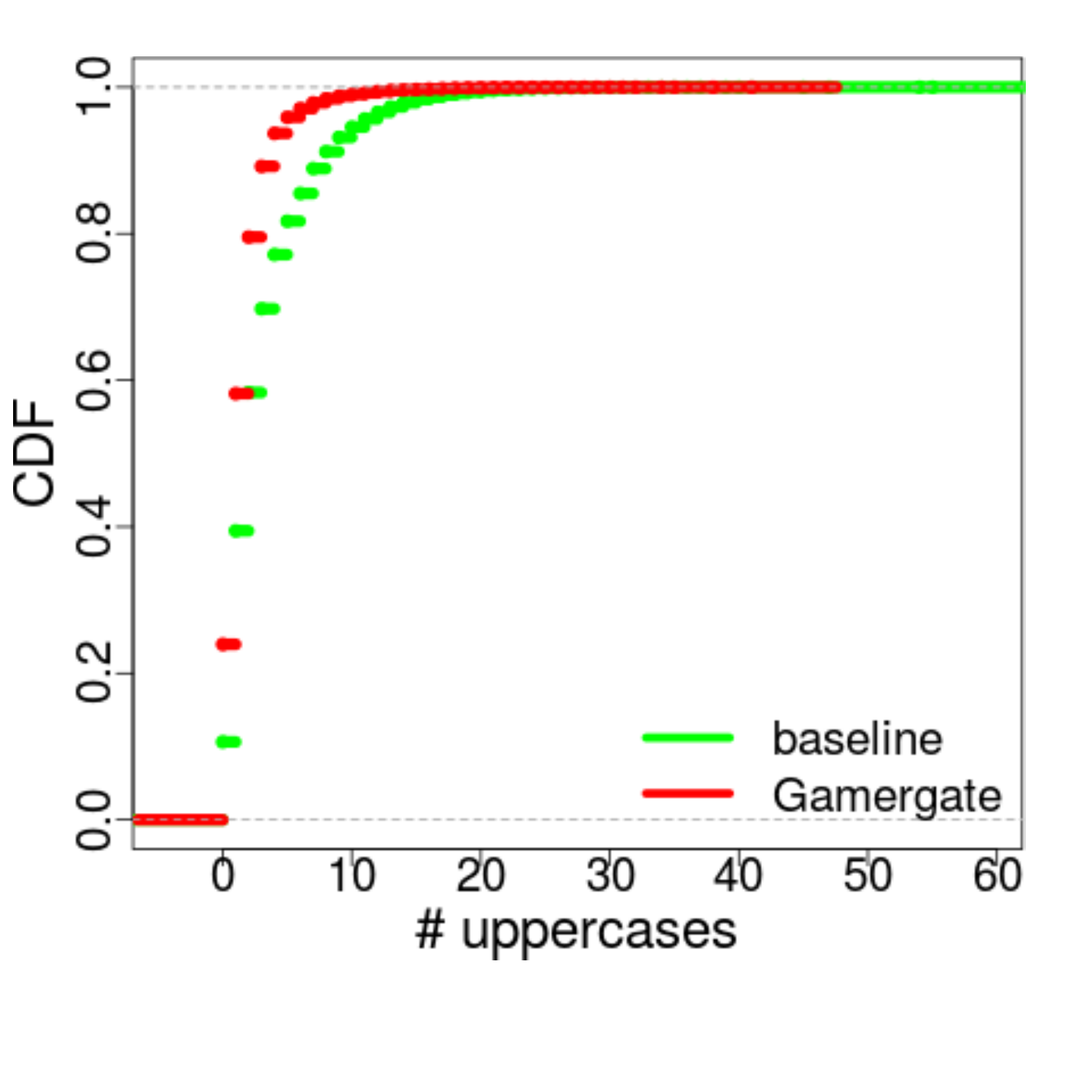}
		\captionsetup{font=scriptsize}
		\vspace{-0.6cm}
		\caption{Uppercases distribution.}
		\label{fig:baseline_hatebase_uppercases}
	\end{subfigure}
	\begin{subfigure}[b]{0.24\textwidth}
		\includegraphics[width=\textwidth]{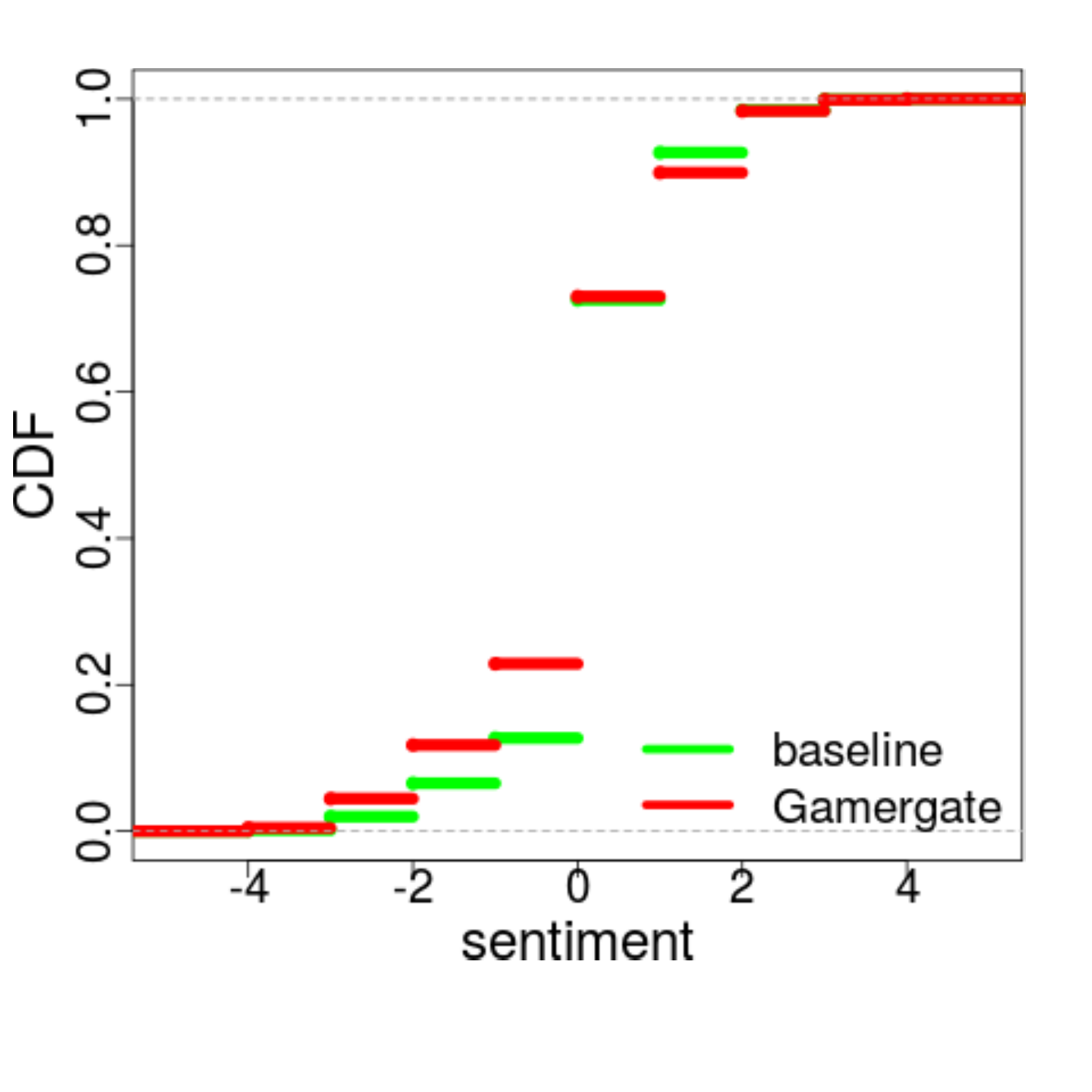}
		\captionsetup{font=scriptsize}
		\vspace{-0.6cm}
		\caption{Sentiment distribution.}
		\label{fig:baseline_hatebase_sentiment}
	\end{subfigure}
	\begin{subfigure}[b]{0.24\textwidth}
		\includegraphics[width=\textwidth]{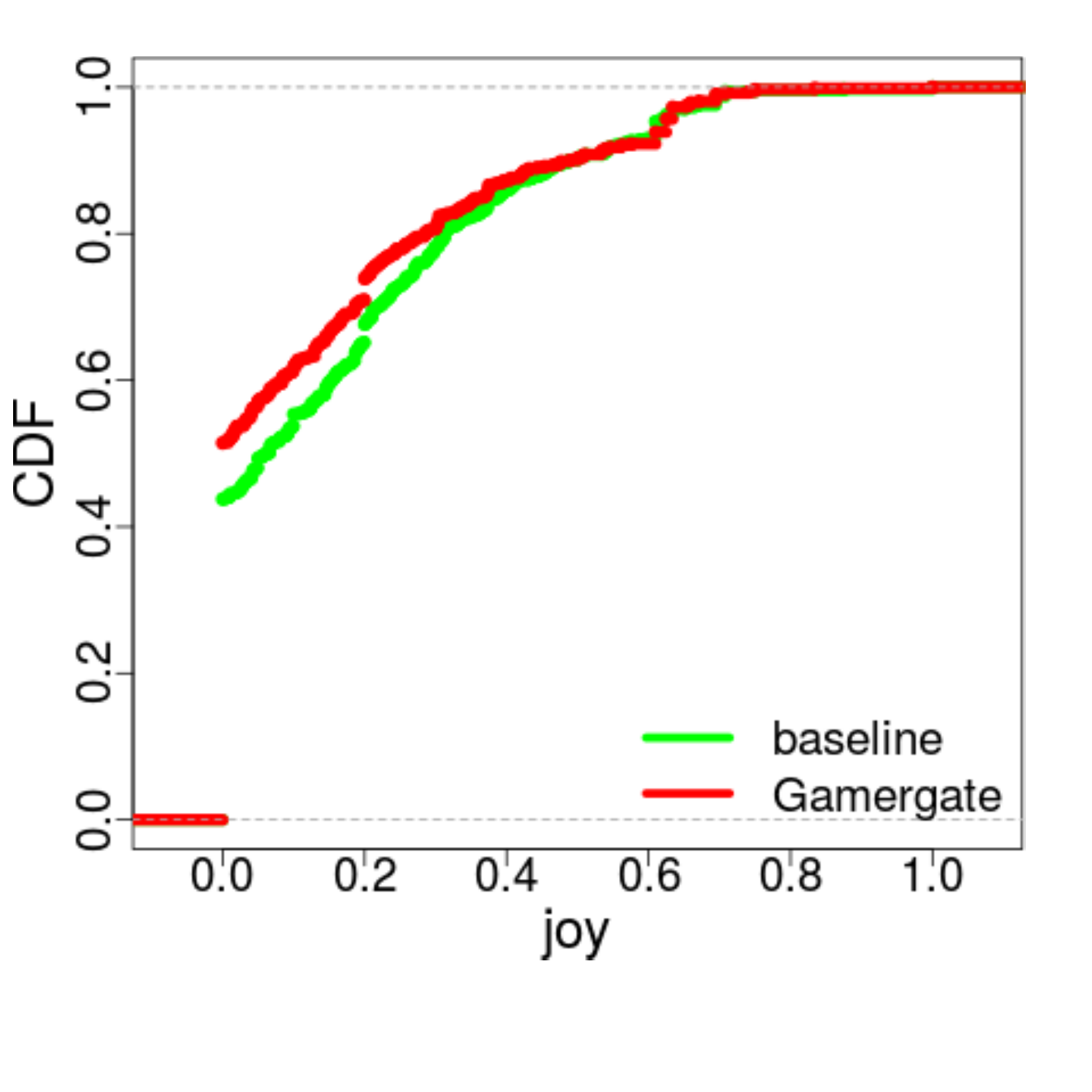}
		\captionsetup{font=scriptsize}
		\vspace{-0.6cm}
		\caption{Joy distribution.}
		\label{fig:baseline_hatebase_joy}
	\end{subfigure}
	\caption{CDF of (a) Emoticons, (b) Uppercases, (c) Sentiment, (d) Joy.}
\end{figure*}

\subsection{How Social are Gamergaters?}

Gamergaters are involved in what we would typically think of as anti-social behavior.
However, this is somewhat at odds with the fact that their activity takes place primarily on social media. 
Aiming to give an idea of how ``social'' Gamergaters are, in Figures~\ref{fig:baseline_hatebase_friends} and~\ref{fig:baseline_hatebase_followers}, we plot the distribution of friends and followers for GG users vs baseline users.
We observe that, perhaps surprisingly, GG users tend to have more friends and followers ($D=0.34$ and $0.39$, $p < 0.01$ for both).
Although this might be somewhat counter-intuitive, the reality is that Gamergate was born on social media, and the controversy appears to be a clear ``us vs. them'' situation.
This leads to easy identification of in-group membership, thus heightening the likelihood of relationship formation.

The ease of in-group membership identification is somewhat different than polarizing issues in the real world where it may be difficult to know a person's views on a polarizing subject, without actually engaging them on the subject.
In fact, people in real life might be unwilling to express their viewpoint because of social consequences.
On the contrary, on social media platforms like Twitter, (pseudo-)anonymity often removes much of the inhibition people feel in the real world, and public timelines can often provide persistent and explicit expression of viewpoints.

\subsection{How Different is Gamergater's Content?}

\descr{Emoticons and Uppercase Tweets.}
Two common ways to express emotion in social media are emoticons and ``shouting'' by using all capital letters.
Based on the nature of Gamergate, we would expect  a relatively low number of emoticon usage, but many tweets that would be shouting in all uppercase letters.
However, as we can see in Figures~\ref{fig:baseline_hatebase_emoticons} and~\ref{fig:baseline_hatebase_uppercases}, which plot the CDF of emoticon usage and all uppercase tweets, respectively, this is not the case.
GG and random users tend to use emoticons at about the same rate (we are unable to reject the null hypothesis with $D=0.028$ and $p = 0.96$).
However, GG users tend to use all uppercase \emph{less} often ($D=0.212$, $p < 0.01$).
As mentioned, GG users are savvy Twitter users, and generally speaking, shouting tends to be ignored.
Thus, one explanation for this behavior is that GG users avoid such a simple ``tell'' as posting in all uppercase, to ensure their message is not so easily dismissed.

\descr{Sentiment.} In Figure~\ref{fig:baseline_hatebase_sentiment}, we plot the CDF of sentiment of tweets.
In both cases (GG and baseline) around 25\% of tweets are positive.
However, GG users post tweets with a generally more negative sentiment ($D=0.101$, $p < 0.01$).
In particular, around 25\% of GG tweets are negative compared to only around 15\% for baseline users.
This observation aligns with the fact that the GG dataset contains a large proportion of offensive posts.

We also compare the offensiveness score of tweets according to Hatebase, a crowdsourced list of hate words.
Each word included in HB is scored on a [0, 100] scale, which indicates how hateful it is.
Though the difference is slight, GG users use more hate words than random users ($D=0.006$, $p < 0.01$). The mean and standard deviation values for HB score are $0.06$ and $2.16$ for the baseline users, while for the GG users they are $0.25$ and $3.55$, respectively.

Finally, based on~\cite{emotions2013}, we extract sentiment values for $6$ different emotions: anger, disgust, fear, joy, sadness, and surprise.
We note that of these, the 2-sample KS test is unable to reject the null hypothesis \emph{except} for joy, as shown in Figure~\ref{fig:baseline_hatebase_joy} ($D=0.089$, $p < 0.01$).
This is particularly interesting because it contradicts the narrative that Gamergaters are posting virulent content out of anger. Instead, GG users are less joyful, and this is a subtle but important difference: they are not necessarily angry, but they are apparently not happy.

\subsection{Are\hspace{-0.01cm} Gamergaters\hspace{-0.01cm} Suspended\hspace{-0.01cm} More\hspace{-0.01cm} Often?}

A Twitter user can be in one of the following three statuses: \emph{active}, \emph{deleted}, or \emph{suspended}.
Typically, Twitter suspends an account (temporarily or even permanently, in some cases) if it has been hijacked/compromised, is considered spam/fake, or if it is \emph{abusive}.\footnote{ \url{https://support.twitter.com/articles/15790}}
A user account is \emph{deleted} if the user-owner of the account deactivates their account.
In the following, we examine the differences among these three statuses with respect to GG and baseline users.

\begin{table}[!t]
\centering
\begin{tabular}{@{}llll@{}}
\toprule
            & active & deleted & suspended \\ \midrule
Baseline    & 67\%   & 13\%    & 20\%      \\
Gamergate & 86\%   & 5\%     & 9\%       \\ \bottomrule
\end{tabular}
\vspace{-0.1cm}
\caption{Status distribution.}
\label{tbl:status}
\end{table}

To examine these differences, we focus on a sample of 33k users from both the GG and baseline datasets.
From Table~\ref{tbl:status}, we observe that, in both cases, users tend to be suspended more often than deleting their accounts by choice.
However, baseline users are more prone to be suspended (20\%) or delete their accounts (13\%) than GG users (9\% and 5\%, respectively).
This seems to be in line with the behavior observed in Figure~\ref{fig:baseline_hatebase_age}, which shows that GG users have been in the platform for a longer period of time; somewhat surprising given their exhibited behavior.
Indeed, a small portion of these users may be spammers who are difficult to detect and filter out.
Nevertheless, Twitter has made significant efforts to address spam and we suspect there is a higher presence of such accounts in the baseline dataset, since the GG dataset is very much focused around a somewhat niche topic.

These efforts are less apparent when it comes to the bullying and aggressive behavior phenomena observed on Twitter in general~\cite{salon,guardiantrolls}, and in our study of Gamergate users in particular.
However, recently, Twitter has increased its efforts to combat the existing harassment cases, for instance, by preventing suspended users from creating new accounts~\cite{CNNtech}, or temporarily limiting users for abusive behavior~\cite{independent}.
Such efforts constitute initial steps to deal with the ongoing war among the abusers, their victims, and online bystanders.

\section{Conclusion}\label{sec:discussion}

This paper presented a first-of-its-kind effort to quantitatively analyze the Gamergate controversy.
We collected 1.6M tweets from 340k unique users using a generic methodology (which can also be used for other platforms and other case studies).
Although focused on a narrow slice of time, we found that, in general, users tweeting about Gamergate  appear to be Twitter savvy and quite engaged with the platform.
They produce more tweets than random users, and have more friends and followers as well.
Surprisingly, we observed that, while expressing more negative sentiment overall, these users only differed significantly from random users with respect to joy.
Finally, we looked at account suspension, finding that Gamergate users are less likely to be suspended due to the inherent difficulties in detecting and combating online harassment activities.

While we believe our work contributes to understanding large-scale online harassment, it is only a start.
As part of future work, we plan to perform a more in-depth study of Gamergate, focusing on how it evolved over time. 
Overall, we argue that a deeper understanding of how online harassment campaigns function can enable our community to better address them and propose detection tools as well as mitigation strategies.

\descr{Acknowledgements.} This work has been funded by the European Commission as part of the ENCASE project  (H2020-MSCA-RISE), under GA number 691025.

\bibliographystyle{abbrv}

\end{document}